\documentclass[conference]{IEEEtran}
\IEEEoverridecommandlockouts

\usepackage{multirow}
\usepackage{pifont}
\usepackage{cite}
\usepackage{amsmath,amssymb,amsfonts}
\usepackage{algorithmic}
\usepackage{graphicx}
\usepackage{textcomp}
\usepackage{xcolor}
\usepackage{soul}
\usepackage{wrapfig}
\usepackage{tabularx}
\usepackage{float}

\def\BibTeX{{\rm B\kern-.05em{\sc i\kern-.025em b}\kern-.08em
    T\kern-.1667em\lower.7ex\hbox{E}\kern-.125emX}}

\newif\ifcomment

\commenttrue

\ifcomment
\definecolor{stelios_colour}{RGB}{144, 238, 144}
\newcommand{\stelios}[1]{\sethlcolor{stelios_colour}\hl{[\textbf{Stelios:} #1]}}
\definecolor{giannis_colour}{RGB}{191, 232, 255}
\newcommand{\giannis}[1]{\sethlcolor{giannis_colour}\hl{[\textbf{Giannis:} #1]}}
\else
\newcommand{\stelios}[1]{}
\newcommand{\giannis}[1]{}
\fi

\renewcommand\IEEEkeywordsname{Keywords}

\begin{document}

\title{Dynamic Temporal Positional Encodings for Early Intrusion Detection in IoT\\
    \thanks{%
        \begin{wrapfigure}[3]{l}{0.202\columnwidth}
            \vspace{-1.1\baselineskip}
            \centering
            \includegraphics[width=0.235\columnwidth]{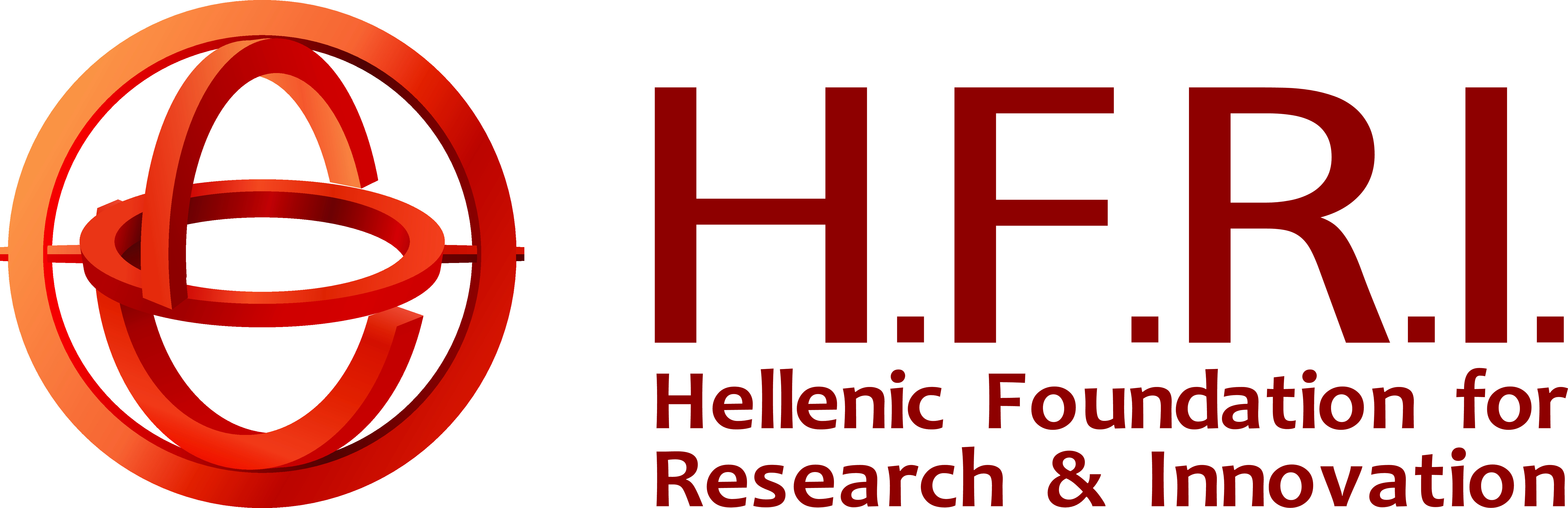}
        \end{wrapfigure}%
        This research work was supported by the Hellenic Foundation for Research and Innovation (HFRI) under the 3rd Call for HFRI PhD Fellowships (Fellowship Number: 5578).
    }
}

\author{
\IEEEauthorblockN{Ioannis Panopoulos}
\IEEEauthorblockA{\textit{School of Electrical and Computer} \\
\textit{Engineering, NTUA, Athens, Greece} \\
ioannispanop@mail.ntua.gr}
\and
\IEEEauthorblockN{Maria-Lamprini A. Bartsioka}
\IEEEauthorblockA{\textit{School of Electrical and Computer} \\
\textit{Engineering, NTUA, Athens, Greece} \\
bartsiokamarilina@mail.ntua.gr}
\and
\IEEEauthorblockN{Sokratis Nikolaidis}
\IEEEauthorblockA{\textit{School of Electrical and Computer} \\
\textit{Engineering, NTUA, Athens, Greece} \\
sokratisnikolaidis@mail.ntua.gr}
\and
\IEEEauthorblockN{Stylianos I. Venieris}
\IEEEauthorblockA{\textit{Samsung AI Center} \\
Cambridge, UK \\
s.venieris@samsung.com}
\and
\IEEEauthorblockN{Dimitra I. Kaklamani}
\IEEEauthorblockA{\textit{School of Electrical and Computer} \\
\textit{Engineering, NTUA, Athens, Greece} \\
dkaklam@mail.ntua.gr}
\and
\IEEEauthorblockN{Iakovos S. Venieris}
\IEEEauthorblockA{\textit{School of Electrical and Computer} \\
\textit{Engineering, NTUA, Athens, Greece} \\
venieris@cs.ece.ntua.gr}
}

\maketitle

\begin{abstract}
The rapid expansion of the Internet of Things (IoT) has introduced significant security challenges, necessitating efficient and adaptive Intrusion Detection Systems (IDS). Traditional IDS models often overlook the temporal characteristics of network traffic, limiting their effectiveness in early threat detection. We propose a Transformer-based Early Intrusion Detection System (EIDS) that incorporates dynamic temporal positional encodings to enhance detection accuracy while maintaining computational efficiency. By leveraging network flow timestamps, our approach captures both sequence structure and timing irregularities indicative of malicious behaviour. Additionally, we introduce a data augmentation pipeline to improve model robustness. Evaluated on the CICIoT2023 dataset, our method outperforms existing models in both accuracy and earliness. We further demonstrate its real-time feasibility on resource-constrained IoT devices, achieving low-latency inference and minimal memory footprint.
\end{abstract}


\section{Introduction}
\label{sec:intro}

The Internet of Things (IoT) enables smart devices to exchange data in real time across various domains, such as smart homes, healthcare, and industrial automation. These systems integrate the physical and digital worlds, generating diverse data types while often operating autonomously. However, their widespread adoption introduces critical security risks due to limited computational resources, weak authentication mechanisms, and exposure to untrusted networks. As a result, IoT devices become prime targets for cyberattacks, including malware, DDoS, and MitM exploits. Ensuring IoT security is crucial~\cite{omolara22compsec}, as breaches can lead to data leaks, service disruptions, and even physical harm in safety-critical applications.

Intrusion Detection Systems (IDS) help mitigate security risks by monitoring network traffic or host activity to detect malicious behaviour. They are broadly categorised into host-based IDS (HIDS), which analyse system logs and resource usage on individual devices, and network-based IDS (NIDS), which inspect packet flows across a network. In recent years, machine learning (ML) and deep learning (DL) have significantly improved IDS capabilities, surpassing traditional rule-based systems by detecting complex attack patterns and zero-day threats. This is particularly important for IoT security, where network traffic is highly dynamic. Deep learning models such as CNNs~\cite{mohammadpour22apllied}, RNNs~\cite{ullah22acccess}, Transformers~\cite{manocchio24expert}, and hybrid architectures~\cite{saiyedand24tmlcn} have been increasingly used for network flow analysis, enabling scalable and real-time threat detection.

A key challenge in network security is real-time threat detection to minimise response times and attack impact. Early Intrusion Detection Systems (EIDS)~\cite{lopez19nca} classify intrusions as early as possible within a session. While Transformers excel in sequential data processing, especially in natural language processing (NLP), network traffic differs as a time series. In this context, packet arrival times provide critical contextual information that is often overlooked by existing IDS. This work proposes a host-based Transformer EIDS for IoT, integrating time-aware positional encodings to capture both sequence structure and temporal dynamics. The system aims for fast, accurate threat detection without sacrificing efficiency. Our key contributions include:
\begin{itemize}
    \item A host-based early intrusion detection system with novel timestamp-based positional encoding mechanisms for rapid and lightweight attack detection.
    \item An augmentation pipeline for network traffic data that improves model robustness and generalisation.
    \item An evaluation of real-world feasibility, demonstrating the system’s efficiency on resource-constrained IoT devices.
\end{itemize}



\section{Related Work \& Background}
\label{sec:related}

\subsection{Intrusion Detection Systems}
\label{ssec:related_ids}

Deep learning-based IDS have gained considerable attention, with recent studies exploring models that process raw network traffic directly, eliminating the need for handcrafted features, which can be computationally expensive and time-consuming. These approaches utilise architectures such as CNNs~\cite{tekerek21compsec}, RNNs~\cite{shahhosseini22jnsm}, Transformers~\cite{han23compsec}, and hybrid models~\cite{zhang19access}, among others. By extracting hierarchical and sequential representations from packet streams, these models improve detection accuracy and adaptability to emerging attack patterns.

Despite these advancements, only a limited number of studies~\cite{ahmad24icstw, islam23cloudcom, ahmad23icstw, ahmad22icstw} have trained models on variable-length network flows, a crucial aspect of early threat detection. However, these works overlook the temporal characteristics of packet flows, which are essential for accurately identifying attacker behaviour. Our approach bridges this gap by explicitly integrating packet timestamps into the detection process using novel dynamic temporal positional encodings for Transformers.

\subsection{Transformer Positional Encodings}
\label{ssec:related_traditional_pes}

In Transformer-based architectures, positional encodings play a crucial role in enabling the model to process sequential data effectively. Since Transformers lack an inherent notion of order, these encodings provide information about the relative or absolute positions of elements within a sequence.

\noindent
\textbf{Sinusoidal Positional Encoding.}
Originally introduced in the Transformer architecture~\cite{vaswani17nips}, sinusoidal positional encodings are incorporated into the input embeddings to encode position-specific information
without requiring learned embeddings.
%
Given a sequence of length \(n\), the sinusoidal positional encoding assigns each position \( p \in \{0, 1, 2, \dots, n-1\} \) a vector of dimension \(d_{\text{m}}\) using the following formulas:
\begin{align*}
    PE(p, 2i) &= \sin \left( p / 10000^{2i/d_{\text{m}}} \right) \\
    PE(p, 2i+1) &= \cos \left( p / 10000^{2i/d_{\text{m}}} \right)
\end{align*}
where \(i = 0, 1, \dots, d_\text{m}/2-1 \) indexes the sine and cosine components within the encoding dimension.

\noindent
\textbf{Fourier-Based Positional Encoding.}
Fourier-based positional encodings~\cite{li21nips} extend the idea of sinusoidal encodings by leveraging a more general Fourier feature mapping. Instead of using a fixed base, these encodings are derived from a learnable frequency basis that enables richer and more flexible positional representations. 
%
The Fourier positional encoding at position \(p\) is defined as:
\begin{align*}
    PE(p, 2i) &= \sin ( 2\pi f_i p ) \\
    PE(p, 2i+1) &= \cos ( 2\pi f_i p ) 
\end{align*}
where \(f_i\) is a learnable frequency parameter associated with the \(i\)-th sine-cosine pair in the encoding.

\noindent
\textbf{Rotary Positional Encoding.}
Another widely adopted method 
is the rotary positional encoding (RoPE)~\cite{su24neurocomputing}, which integrates positional information directly into the self-attention mechanism. This approach enables the model to encode relative positional dependencies naturally, enhancing its ability to capture long-range relationships in sequential data.
%
Standard RoPE applies a rotation matrix to both the query and key vectors in the self-attention mechanism. Let \(\mathbf{x} = \left( x_0, x_1, \dots, x_{d_\text{m}-1}\right)\) represent a packet embedding.
%
%
For each index \(i\), the corresponding subvector, consisting of two consecutive embedding elements \( \left( x_{2i} , x_{2i+1} \right) \),
is transformed as follows:
\begin{equation*}
    \begin{bmatrix} x_{2i}^{\text{rot}} \\ x_{2i+1}^{\text{rot}} \end{bmatrix} =
    \begin{bmatrix} \cos(p\theta_i) & -\sin(p\theta_i) \\ \sin(p\theta_i) & \cos(p\theta_i) \end{bmatrix}
    \begin{bmatrix} x_{2i} \\ x_{2i+1} \end{bmatrix}
\end{equation*}
where 
\(\theta_i\) is the rotational angle defined as \( \theta_i = 10000^{-i/d_\text{m}} \).


Some recent studies have explored time-aware positional encodings by leveraging timestamps in sequential data, but these methods often introduce significant computational overhead~\cite{rashed22recsys, ji25appint}, making them impractical for resource-constrained environments.
Additionally, existing works focus solely on extending sinusoidal positional encodings~\cite{ryu23sera, sharma23icmla}, limiting their applicability to alternative encoding strategies that may be better suited for irregularly spaced packets. To our knowledge, this is the first study to apply time-aware positional encodings to network traffic analysis and systematically evaluate their impact across three different encoding mechanisms.

\section{Our Approach}
\label{sec:approach}

Host-based intrusion detection systems are particularly effective in identifying attacks that target a single host. These attacks can be classified as either one-to-one, where a single attacker targets a single victim, or many-to-one, where multiple sources coordinate to overwhelm a single victim. This study focuses on one-to-one attacks, as they present a greater challenge for detection due to their stealthy nature.

To effectively detect such attacks, an analysis at the flow level is crucial.
According to RFC 7011~\cite{rfc7011}, a \textit{traffic flow} is a set of packets or frames passing through an observation point over a specified time interval. In the context of our host-based IDS, the observation point is the
IoT device vulnerable to attacks. Each packet within a flow shares common attributes, with one of the most widely accepted definitions being the 5-tuple expression: source and destination IP addresses, source and destination transport layer ports, and the protocol in use.


\begin{figure}[t]
    \centering
    \includegraphics[width=\linewidth,trim={0.45cm 2.3cm 0.45cm 2.3cm},clip]{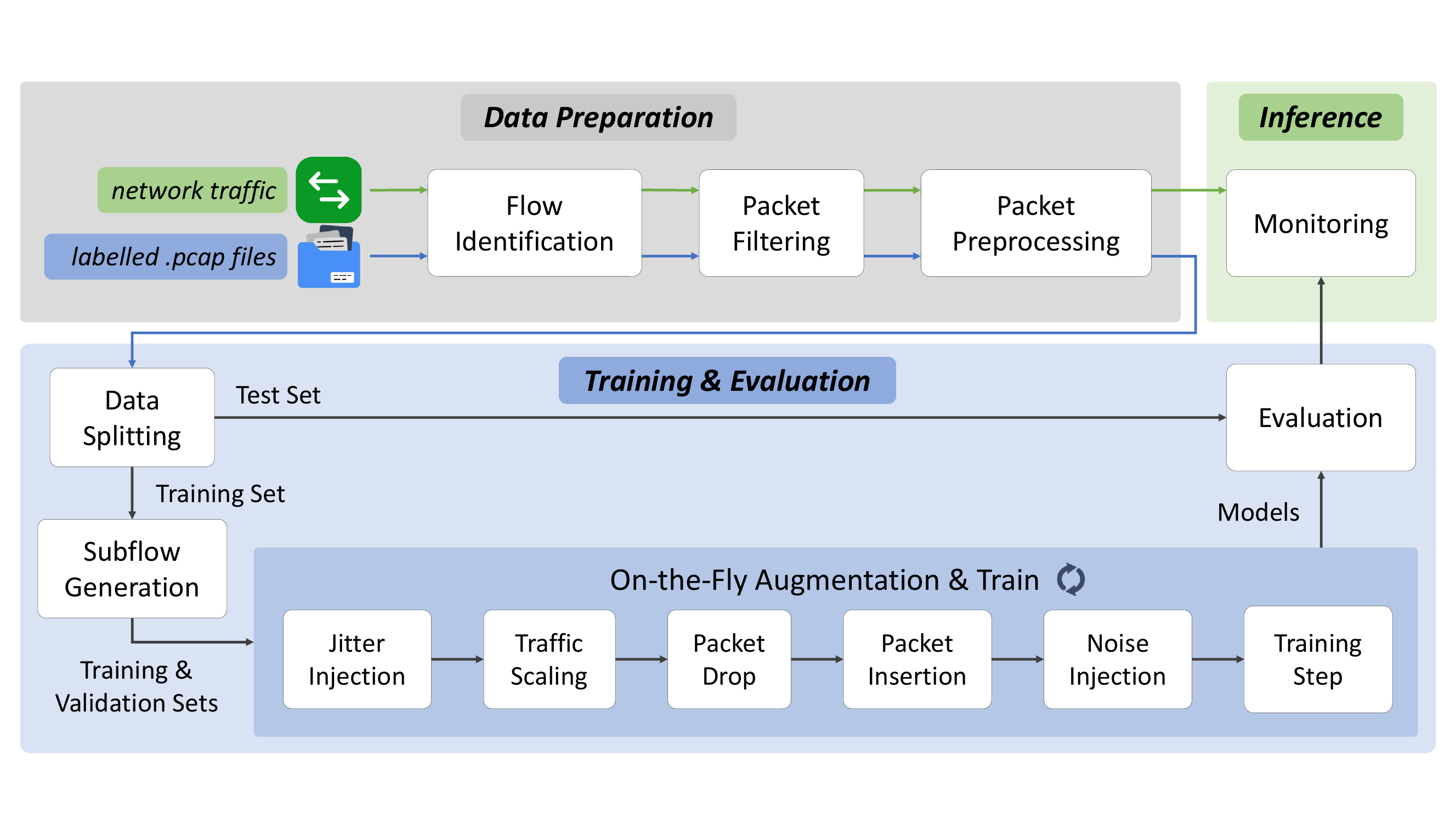}
    \caption{The Three-Stage Architecture of the Proposed EIDS.}
    \label{fig:system}
\end{figure}

Fig.~\ref{fig:system} illustrates the proposed system for early intrusion detection, comprising three main stages.
The \textit{Data Preparation} stage (Section~\ref{ssec:data_prep}), which is common to both training and inference, involves processing raw network traffic and labelled PCAP files 
to ensure that data is structured appropriately for subsequent analysis. In the \textit{Training \& Evaluation} stage (Section~\ref{ssec:training}), the processed data is first split into training and test sets. The training data undergoes On-the-Fly Augmentation, which includes a pipeline of techniques targeted to network data, followed by the training step. The trained models are then evaluated using the test set. Finally, the \textit{Inference} stage employs the trained models for real-time monitoring of network traffic. 


\subsection{Data Preparation}
\label{ssec:data_prep}

Our goal in converting raw network traffic into a format suitable for deep neural network (DNN) training is to maximise computational efficiency and enable rapid processing. To achieve this, we depart from traditional methods that rely on extracting predefined features and instead utilise raw packet bytes as direct input to our model. The following subsections describe the three modules involved in this process.


\subsubsection{Flow Identification}
\label{sssec:flow_id}
The initial step in the proposed system involves maintaining a record of all network flows associated with the host, regardless of whether they originate from network captures 
or real-time network traffic.
%
Formally, a flow \(F\) can be represented as an ordered sequence of packets:
\begin{equation}
    F = \{ P_1, P_2, ..., P_n \}, P_i \in \mathbb{R}^d
    \label{eq:flow_def}
\end{equation}
%
%
where \(P_i\) is the \(i\)-th packet, \(n\) is the length of the flow and \(d\) is the length of a packet in number of bytes. In our system, each flow can have a maximum length \(N\), 
therefore, the number of packets in any given flow satisfies the condition \( 1 \leq n \leq N \).


\subsubsection{Packet Filtering}
\label{sssec:packet_filter}
The second step 
involves isolating network packets relevant to the threats targeted by the IDS. By filtering out irrelevant packets before further analysis, the system reduces computational complexity and directs its focus toward potentially malicious traffic. Certain network protocols, such as HTTP, ARP, and ICMP, are commonly exploited for attacks, making their targeted analysis beneficial for threat detection.

\subsubsection{Packet Preprocessing}
\label{sssec:packet_preprocess}

The final step, packet preprocessing, converts raw network packets into a structured format suitable for deep learning models. It removes irrelevant headers, including the Ethernet header and IP addresses, to prevent potential model overfitting to specific address patterns. Each packet is then truncated or padded to a fixed length, denoted as \(d\) in Equation~\eqref{eq:flow_def}, ensuring uniform input dimensions. Lastly, byte values are normalised to [0,1] to improve training stability~\cite{huang2020pami}. Packet timestamps are also extracted to capture inter-arrival times (IAT), providing crucial temporal context. Stored as a separate vector \(T\), timestamps start at 0 for the first packet, with subsequent values representing the absolute time elapsed since the first packet's arrival.

\subsection{Training}
\label{ssec:training}

The training phase of the proposed system is designed to develop deep learning models capable of identifying one-to-one attacks based on raw network packet data. This phase follows the data preparation stage, utilising the preprocessed packet representations and corresponding timestamp vectors as input.

\subsubsection{Data Splitting}
\label{sssec:data_split}
First, the dataset is divided into training and test sets.
%
Since Transformers process batches of fixed-length sequences, all network flows are padded to a uniform length before training. To achieve this, each flow in the training set is zero-padded to match the maximum flow length \(N\), ensuring consistency. This process also generates attention masks, which guide the Transformer's attention mechanism.


\subsubsection{Augmentation}
\label{sssec:augmentation}

Before each training iteration, augmentation is applied to the training flows to enhance generalisation, robustness, and early detection capabilities. To achieve this, subflows are generated by retaining only the first \(k\) packets of each flow, where \(k\) varies from 1 to the full flow length. This approach trains the model to classify flows based on partial information, rather than relying on the entire sequence. Such early classification is crucial for reducing response times by detecting threats from only a few initial packets.


Additionally, on-the-fly (epoch-wise) augmentation techniques are applied to each (sub)flow to increase dataset diversity and mitigate overfitting. These augmentations are performed independently on each training sample, exposing the model to a broader range of realistic, slightly modified network flows. By incorporating packet timestamps, timestamp-based augmentations further enhance dataset variability. Five augmentation techniques, each targeting specific flow characteristics, are applied sequentially, as illustrated in Fig.~\ref{fig:system}.

\noindent
\textbf{Jitter Injection.}
%
%
%
This technique simulates real-world network jitter by introducing small, random variations in packet arrival times. For each timestamp in a flow, the minimum temporal distance between the previous and next timestamp, denoted as \(t_{\text{min}}\), is computed. A random perturbation is then applied, drawn from the continuous uniform distribution \( \mathcal{U}(-0.7 \cdot t_{\text{min}}, 0.7 \cdot t_{\text{min}}) \), ensuring realistic timing fluctuations.


\noindent
\textbf{Traffic Scaling.}
This method emulates varying network speeds by applying a randomly selected scaling factor from the set \( \{0.5, 0.75, 1.0, 1.25, 1.5\} \) to the inter-packet times. Scaling up simulates slower networks, whereas scaling down mimics high-speed links. This variation helps the model adapt to diverse network conditions, enhancing its ability to generalise across different traffic speeds.

\noindent
\textbf{Packet Drop.}
%
%
This augmentation technique randomly drops a number of packets from each flow. The maximum number of packets that can be dropped depends on the length of the flow, calculated as \( \textit{max\_packets\_to\_drop} = \lfloor 0.25 \cdot n - 0.5 \rfloor \), where \(n\) is the length of the flow. The actual number of packets to drop is drawn from the discrete uniform distribution \( \mathcal{U} \{0, \textit{max\_packets\_to\_drop} \} \).


\noindent
\textbf{Packet Insertion.}
The packet insertion technique randomly adds a number of zero-byte packets into a flow. The maximum number of zero packets to be inserted is based on the flow length and is calculated as \( \textit{max\_packets\_to\_insert} = \lfloor 0.15 \cdot n - 0.5 \rfloor \). The actual number is again drawn from the discrete uniform distribution \( \mathcal{U} \{0, \textit{max\_packets\_to\_insert} \} \).


\noindent
\textbf{Noise Injection.}
The final augmentation method involves adding noise to the packets. For each flow, we choose to modify at most \(\lfloor n/3 \rfloor\) packets, and for each modified packet, at most \( \lfloor d/100 \rfloor\) bytes are altered. The positions of the altered bytes are randomly selected from a discrete uniform distribution, while the noise itself is drawn from a continuous normal distribution with zero mean and a standard deviation of 0.1.

\subsubsection{Early Detection Loss Function}
\label{sssec:loss_fn}
%
%

To enhance early classification performance, we introduce Early Detection Loss (EDL), a custom training loss function that assigns greater penalties to misclassifications occurring in shorter network flows.
During training, the standard cross-entropy loss is first calculated individually for each sample in a batch of size \(b\). The overall batch loss is then computed as a weighted average of these individual losses:
\begin{equation}
    L = \sum_{i=0}^{b-1} w_i CE_i
    \label{eq:loss_fn}
\end{equation}
where \(w_i = e^{-0.1 \cdot n_i} \) represents the weight associated with the \(i\)-th sample based on its flow length \(n_i\), and \(CE_i\) is the corresponding cross-entropy loss. By exponentially reducing the weight as \(n_i\) increases, EDL incentivises the model to optimise for accurate classifications in the earliest possible stage of network flow analysis.
%
%


\subsection{Model}
\label{ssec:model}

\subsubsection{Base Model}
\label{sssec:base_model}

%
%




This subsection outlines the architecture of our base model, which refers to the core Transformer architecture without positional encodings. Our design is based on the standard Transformer framework proposed in~\cite{vaswani17nips}, with key modifications to better suit network flow analysis. Unlike traditional Transformer-based models that require an embedding layer to map discrete tokens to dense representations, our approach processes raw packet byte values directly, treating them as token embeddings. To align the input representation with the model’s internal feature space, a fully connected layer transforms the input dimension, \(d\), into the model's hidden dimension, \(d_\text{m}\).
Additionally, we replace the GELU activation function with ReLU, optimising computational efficiency and training stability.

Since our task is classification, we utilise only the Transformer encoder. After the final Transformer block, we apply global average pooling over the sequence, aggregating packet-level representations into a fixed-length vector. This representation is then passed through a fully connected layer,
followed by a softmax activation function to generate class confidence scores. A detailed overview of the 
final tuned values and model size analysis is provided in Section~\ref{sssec:hyperparam_tuning}.

\subsubsection{Dynamic Temporal Positional Encodings}


Conventional positional encodings rely on predefined position indices, assuming a uniform and fixed spacing between sequence elements. While this assumption is appropriate for structured data such as text, it becomes problematic in network traffic analysis, where packet flows exhibit variable inter-arrival times. As a result, traditional positional encodings may fail to effectively capture the temporal structure of network traffic, limiting their utility in intrusion detection.

To overcome this limitation, we introduce dynamic temporal positional encodings, which replace the predefined position indices \( \mathcal{P} = \{0, 1, 2, \dots, n-1\} \) with the actual packet timestamps in the flow, denoted as \( T = \{t_p\}_{p=0}^{n-1} \). This adaptation allows the model to encode the inherent temporal irregularities in network traffic more accurately. Our approach is applied to sinusoidal, Fourier-based, and rotary positional encodings, resulting in three novel time-aware encoding mechanisms tailored for sequential network data.
Beyond network traffic analysis, the proposed encodings can be extended to various types of time series data, where samples are recorded at uneven intervals. 






\section{Implementation}
\label{sec:impl}

The implementation of our system is structured into three main components: network data processing, model development, and deployment on edge devices. We use Scapy, a Python package for parsing and processing network traffic, to extract relevant packet flows and features from raw PCAP files. The Transformer-based detection model is developed using TensorFlow. To ensure efficient real-time inference on resource-constrained IoT devices, we deploy the trained models on the Raspberry Pi Zero 2 W using LiteRT, a lightweight runtime optimised for deep learning execution. This approach enables low-latency, high-accuracy intrusion detection, making the system suitable for practical IoT security applications.

\section{Evaluation}
\label{sec:exp_setup}

\subsection{Experimental Setup}


\subsubsection{Dataset}
\label{sssec:dataset}
For our experiments, we use the CICIoT2023 dataset~\cite{neto23sensors}, a publicly available benchmark dataset specifically designed for intrusion detection in IoT environments. 
%
%
%
%
Given our system's focus on one-to-one attack detection, we concentrate specifically on web-based threats within the dataset. These attacks exploit vulnerabilities in web applications and communication protocols and often serve as entry points for larger cyberattacks. 
%
%
To evaluate our system, we selected five distinct web-based attack types—SQL Injection, Command Injection, Backdoor Malware, Uploading Attack, and Cross-Site Scripting (XSS)—along with benign traffic, resulting in a six-class classification task (\(c=6\)).

In the context of one-to-one web-based attacks, the conventional 5-tuple flow definition leads to 
multiple short-lived, two-packet flows due to dynamic or randomised ports.
%
%
To address this, we redefine a flow
using a 3-tuple representation \( (\text{IP}_\text{src}, \text{IP}_\text{dst}, \text{Proto}) \), grouping all exchanges between the attacker and victim regardless of port variations. Here, \(\text{IP}_\text{src}\) is the source IP address (initiating communication), \(\text{IP}_\text{dst}\) is the destination IP address, and \(\text{Proto}\) indicates the protocol in use, which in this case is HTTP.


\subsubsection{Hyperparameter Tuning}
\label{sssec:hyperparam_tuning}


The choices for data preparation and base model hyperparameters were guided by the need for fast inference and a lightweight design suitable for real-world network environments. Table~\ref{tab:hyperparams} presents the selected values. In this configuration, the base model comprises only 5,086 trainable parameters, excluding positional encodings. This lightweight architecture facilitates rapid decision-making while maintaining the necessary representational power to distinguish between normal and attack traffic.

\begin{table}[htbp]
  \caption{System Hyperparameters and their Corresponding Values}
  \label{tab:hyperparams}
  \begin{center}
        \begin{tabular}{r | c | r}
        \hline
        \rule{0pt}{8pt} \textbf{Hyperparameter} & \textbf{Symbol} & \textbf{Value} \\
        \hline
        \rule{0pt}{8pt} Packet length & \(d\) & 448 \\
        \rule{0pt}{8pt} Maximum flow length & \(N\) & 30 \\
        \hline
        \rule{0pt}{8pt} Hidden dimension & \(d_\text{m}\) & 8 \\
        \rule{0pt}{8pt} Number of Transformer blocks & \(L\) & 1 \\
        \rule{0pt}{8pt} Number of attention heads & \(h\) & 4 \\
        \rule{0pt}{8pt} Attention head dimension & \(d_\text{h}\) & 8 \\
        \rule{0pt}{8pt} FFN intermediate dimension & \(d_\text{ff}\) & 16 \\
        \rule{0pt}{8pt} Dropout rate & \(p_\text{drop}\) & 0.1 \\
        \rule{0pt}{8pt} Number of output classes & \(c\) & 6 \\
        \hline
        \end{tabular}
    \end{center}
\end{table}

%

%


\subsubsection{Training Process}
The primary challenge in training our system was the limited number of samples. Each web-based attack in CICIoT2023 has only three recorded sessions, resulting in three flows per attack class. To balance the dataset, we extract three benign flows for the benign class. Thus, our initial dataset consists of a total of \(3 \cdot 6 = 18\) samples.

\noindent
\textbf{Ensemble Learning.}
Given the limited training data, we opt for an ensemble learning approach instead of training a single model. Multiple models are trained on different dataset splits to enhance diversity and generalisation. Each split uses two of the three available samples per class for training, with the third reserved for testing. Out of 729 possible splits (\(3^6\)), we select 29 diverse configurations, ultimately retaining only the best-performing models to build a robust ensemble trained on meaningful data partitions.

\noindent
\textbf{Augmentation and Oversampling.}
Each class in our dataset contains flows exceeding \(N=30\) packets, yielding 60 training samples per class after subflow generation. To address the small dataset size, we apply deterministic oversampling as an additional augmentation step, duplicating each sample five times, resulting in 300 samples per class. This fixed duplication is effective, as random augmentations introduced at each training epoch per sample (see Section~\ref{sssec:augmentation}) maintain variability and prevent overfitting.

\noindent
\textbf{Training Configuration.}
%
For model optimisation, we employ the Adam optimiser with a fixed learning rate of 0.0002. Training follows the Early Detection loss function described in Equation~\eqref{eq:loss_fn} and is performed with a batch size of 4 samples.


\subsubsection{Evaluation Metrics}
To assess the effectiveness of our system, we design an evaluation process that closely simulates its real-world deployment scenario (\textit{Inference} stage in Fig.~\ref{fig:system}). 


\noindent
\textbf{Confidence-Based Performance Metrics.}
To assess classification performance, we employ a set of confidence-based metrics. Given a confidence threshold \(\tau\) applied to the top-1 softmax score, we evaluate how quickly the system reaches a confident decision. The process begins with the first packet of each test flow, gradually adding packets until the model's confidence exceeds \(\tau\). If the threshold is not met after processing all \(N\) packets, classification is based on the full sequence.

A key metric is \textit{Earliness}, which measures the number of packets needed before the model reaches the confidence threshold and a correct prediction is made.
At the threshold point, we also compute \textit{Top-1 Accuracy},
\textit{False Negative Rate (FNR)},
and \textit{False Alarm Rate (FAR)}.
A high FNR poses a security risk due to undetected attacks, while minimising FAR prevents unnecessary alerts.
%
Lastly, we utilise the Early Risk Detection Error (ERDE)~\cite{fernandez18nca}, a metric that evaluates both the correctness of the model and the delay in reaching a decision. 
%

\noindent
\textbf{Resource-Constrained Deployment Evaluation.}
%
Beyond classification performance, we assess the feasibility of deploying our system on resource-constrained edge devices, ensuring its practicality for real-world IoT applications. To this end, we implement the system on a Raspberry Pi Zero 2 W, a compact, low-power embedded device featuring a 1GHz quad-core 64-bit ARM Cortex-A53 processor and 512MB of RAM, making it representative of typical IoT hardware limitations.


%



\begin{figure}[t]
    \centering
    \includegraphics[width=0.95\linewidth]{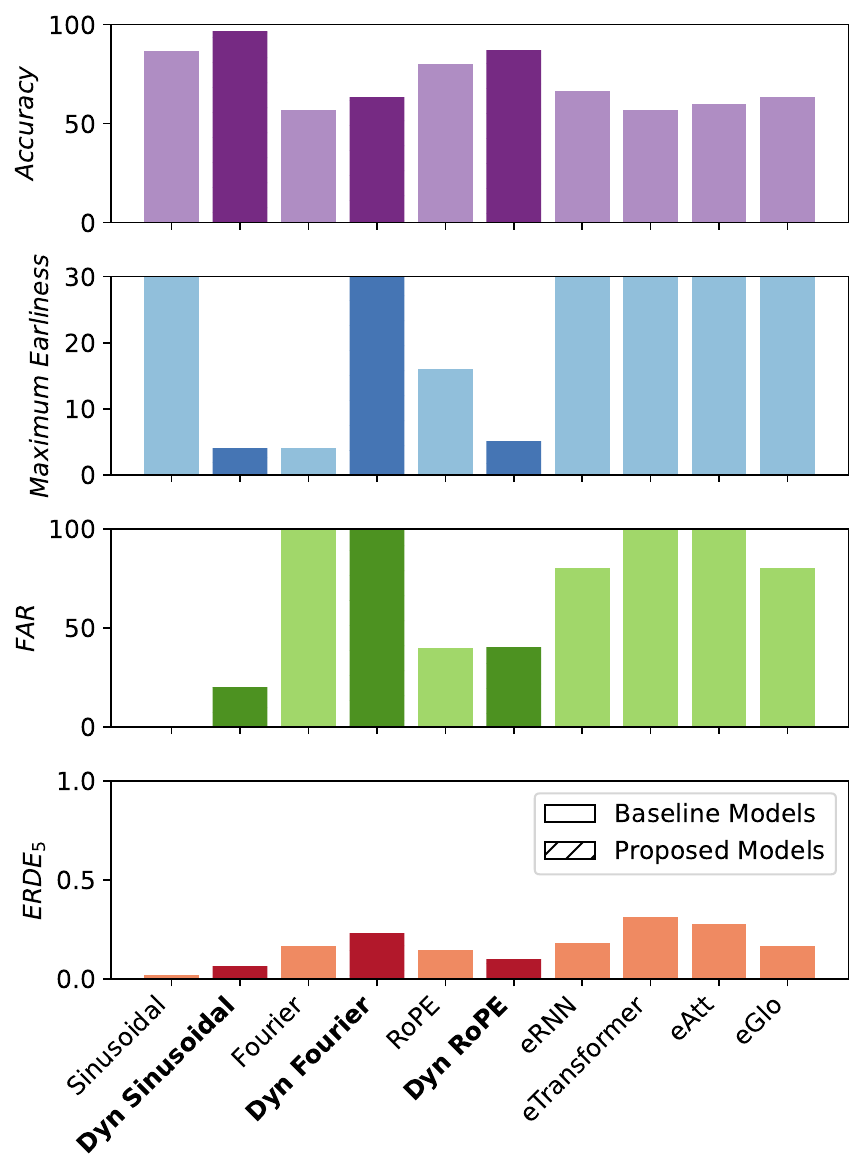}
    \caption{Confidence-Based Evaluation.}
    \label{fig:conf_based}
\end{figure}

\subsection{Results}

%
%
We evaluate our approach against several well-established baselines and relevant prior works. Specifically, we compare it to the non-dynamic variants of sinusoidal, Fourier-based, and rotary positional encodings, which introduce minimal parameter cost—most adding only 0–4 parameters, except for the sinusoidal encoding, which contributes \(d_\text{m} \cdot N = 240\).
%
%
Furthermore, drawing from the related work discussed in Section~\ref{sec:related}, we benchmark our system against four neural network architectures proposed for EIDS in~\cite{ahmad24icstw, islam23cloudcom, ahmad23icstw, ahmad22icstw}, referred to as eRNN, eTransformer, eAtt, and eGlo.

\subsubsection{Earliness \& Accuracy Metrics}

Fig.~\ref{fig:conf_based} presents the performance of the evaluated five-model ensemble systems at a 99\% confidence threshold across all metrics, except FNR, as it remains 0 for all systems except RoPE (4\%) and eTransformer (8\%). The maximum earliness values correspond to the highest accuracy achieved by each system, as accuracy is considered the most critical metric. Lower earliness could be attained for all models if a slight reduction in accuracy were acceptable.

The results demonstrate that the proposed dynamic encodings consistently outperform both their non-dynamic counterparts and baseline models from related work. While some models achieve similar accuracy, they often do so at the cost of increased earliness, making them less suitable for real-time intrusion detection. Among the evaluated configurations, the ensemble incorporating the dynamic sinusoidal positional encoding achieves the best trade-off between accuracy and earliness. Specifically, it reaches an accuracy of 96.67\%, a maximum earliness of 4 packets, a false negative rate of 0\%, a false alarm rate of 20\%, and an ERDE$_5$ score of 0.0656, highlighting its effectiveness in early intrusion detection.


\begin{figure}[b]
    \centering
    \includegraphics[width=0.95\linewidth,trim={9.4cm 1.55cm 8.7cm 2.15cm},clip]{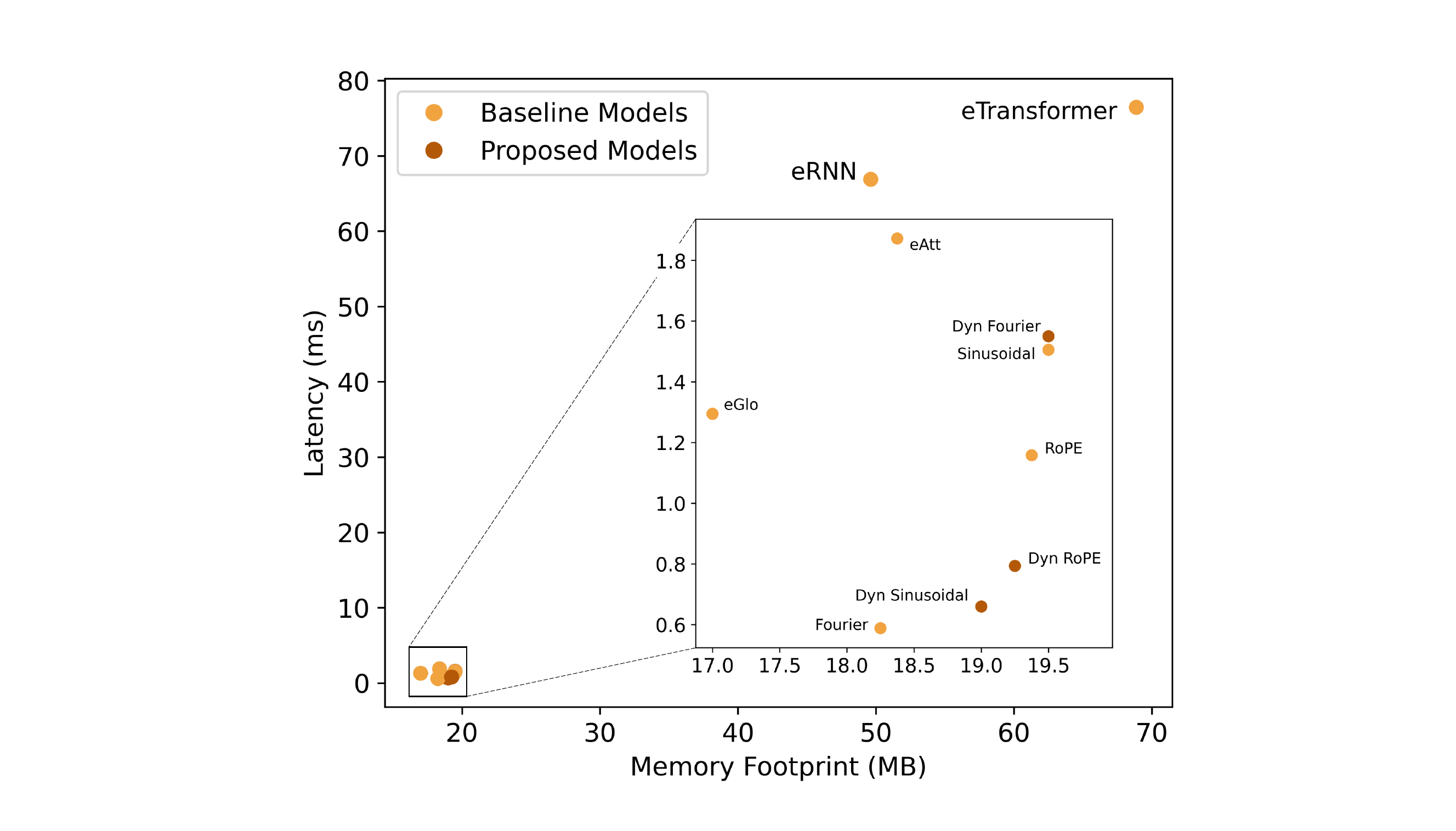}
    \caption{Latency vs. Memory Footprint.}
    \label{fig:latency_mf}
\end{figure}

\subsubsection{Latency \& Memory Footprint}

%

Fig.~\ref{fig:latency_mf} depicts the trade-off between latency and memory footprint when deploying the evaluated systems on the Raspberry Pi Zero 2 W. Compared to baseline models, the proposed dynamic encodings achieve a favourable balance between computational efficiency and resource usage. Notably, all three proposed systems maintain a latency of under 2 ms while keeping their memory footprint below 20 MB. The zoomed-in section of the figure further highlights their efficiency, demonstrating that dynamic sinusoidal and dynamic RoPE encodings, which achieve high accuracy, also exhibit lower latency than most tested configurations, reinforcing their suitability for real-world deployment.










\section{Conclusion}
\label{sec:conclusion}


This work introduces a Transformer-based Early Intrusion Detection System (EIDS) that enhances IoT threat detection using dynamic temporal positional encodings. By incorporating packet timestamps, the model improves detection accuracy and earliness. Evaluation on CICIoT2023 shows that dynamic encodings outperform traditional methods, achieving high accuracy, lower false alarm rates, and faster classification. Real-world tests confirm the system's feasibility on resource-constrained edge devices, ensuring low latency and minimal memory usage. Future directions include expanding evaluation to more datasets, investigating multi-device intrusion detection, and introducing quantisation.


\bibliographystyle{IEEEtran}
{\footnotesize
\bibliography{references}
}

\end{document}